\documentclass[conference]{IEEEtran}
\usepackage{cite}
\usepackage{amsmath,amssymb,amsfonts}
\usepackage{algorithmic}
\usepackage{graphicx}
\usepackage{textcomp}
\usepackage{xcolor}
\usepackage{soul}
\usepackage{url}
\usepackage{multirow}
\usepackage{hyperref}
\usepackage{booktabs}
\usepackage{fancyhdr}
\def\BibTeX{{\rm B\kern-.05em{\sc i\kern-.025em b}\kern-.08em
    T\kern-.1667em\lower.7ex\hbox{E}\kern-.125emX}}
    
\fancypagestyle{FirstPage}{
\lhead{\small{\copyright  2022 IEEE. Personal use of this material is permitted. Permission from IEEE must be obtained for all other uses, in any current or future media, including reprinting/republishing this material for advertising or promotional purposes, creating new collective works, for resale or redistribution to servers or lists, or reuse of any copyrighted component of this work in other works.} }
}

\begin{document}

\title{Device {I}nteroperability for {L}earned {I}mage {C}ompression with {W}eights and {A}ctivations {Q}uantization \\
  {\normalfont\large 
    Esin Koyuncu\IEEEauthorrefmark{1}\IEEEauthorrefmark{2}, Timofey Solovyev\IEEEauthorrefmark{2}, Elena Alshina\IEEEauthorrefmark{2}, Andr\'{e} Kaup\IEEEauthorrefmark{1}%
  }\\[-1.5ex]
}

\author{
    \IEEEauthorblockA{\IEEEauthorrefmark{1}\textit{Multimedia Communications and Signal Processing}\\
    \textit{Friedrich-Alexander-Universit\"{a}t Erlangen-N\"{u}rnberg} \\
    Erlangen, Germany \\
    \{andre.kaup, esin.koyuncu\}@fau.de
    }
\and
    \IEEEauthorblockA{\IEEEauthorrefmark{2}\textit{Audiovisual Laboratory, Munich Research Center} \\
    \textit{Huawei Technologies} \\
        Munich, Germany \\
        \{elena.alshina, solovyev.timofey\}@huawei.com
    }
}
\renewcommand{\headrulewidth}{0pt}

\maketitle

\begin{abstract}
Learning-based image compression has improved to a level where it can outperform traditional image codecs such as HEVC and VVC in terms of coding performance. In addition to good compression performance, device interoperability is essential for a compression codec to be deployed, i.e., encoding and decoding on different CPUs or GPUs should be error-free and with negligible performance reduction. In this paper, we present a method to solve the device interoperability problem of a state-of-the-art image compression network. We implement quantization to entropy networks which output entropy parameters. We suggest a simple method which can ensure cross-platform encoding and decoding, and can be implemented quickly with minor performance deviation, of 0.3\% BD-rate, from floating point model results.
\end{abstract}


\section{Introduction}

\thispagestyle{FirstPage}

Image compression methods can be categorized into two approaches: traditional codecs such as HEVC \cite{hevc_2012overview} and the state-of-the-art VVC \cite{vvc}, and learning-based codecs such as \cite{balle2018variational}, \cite{cheng2020learned} and \cite{minnen2018joint}. These learning-based compression networks consist of two main levels, an analysis-synthesis part, which maps an image to a latent representation; and a hyperprior part, which learns a prior distribution of the latent representation for entropy coding. Prior distributions are learned with different models in previous research, such as zero-mean Gaussian model in \cite{balle2018variational}, Gaussian model with non-zero mean such as in \cite{minnen2018joint}, or Gaussian mixture model as in \cite{cheng2020learned}. Also, a more complex entropy parameter learning approach is suggested in \cite{minnen2018joint} to learn Gaussian parameters jointly, by a hyperprior and context model.

\begin{figure}[btp]
\centerline{\includegraphics[height=2.95cm]{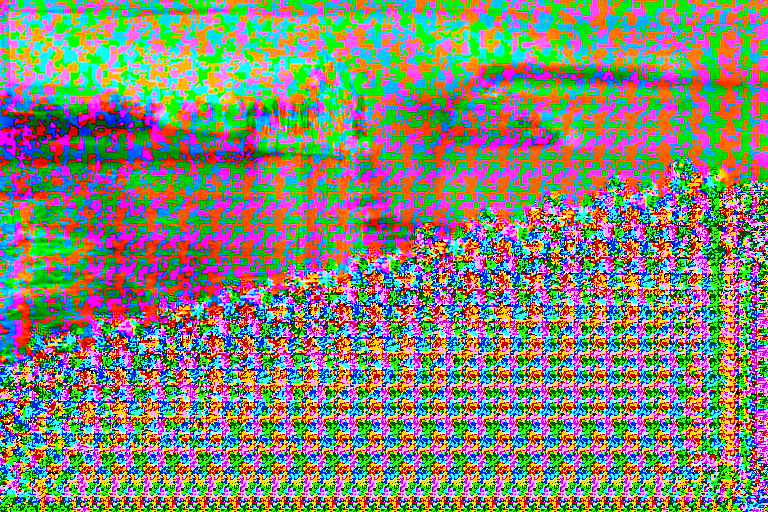}\includegraphics[height=2.95cm]{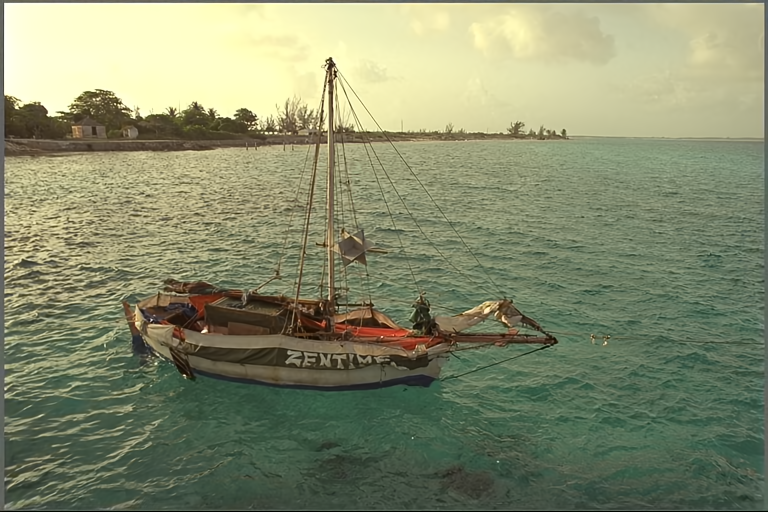}}
\caption{Reconstructed images of test image 6 from decoder when encoder is run on GPU and decoder is run on CPU. On the left: floating point model (Ours-float). On the right: model with integerized entropy (Ours-int16-entropy). 
}
\label{fig:rec_image}
\end{figure}

For better learning of features and representations, floating point arithmetic is used during training of these networks. During evaluation,  we need priors to be calculated correctly at encoder and decoder side to achieve error-free decoding. Since calculation of floating point arithmetic on different devices is deviating, priors calculated at encoder and decoder side can be slightly different from each other. Even a small deviation causes an error in entropy decoding, leading to visible errors in reconstructed images at the decoder as illustrated in Fig. \ref{fig:rec_image}. 

To our knowledge, \cite{balle2018integer} is the first work addressing the learned image compression decoder failure problem on different platforms. In their work, to achieve determinism on different platforms, Ball\'{e} \textit{et al.} propose the solution of converting networks, weights and activations, to integer, and discretizing the entropy parameters. A look up table (LUT) is used to calculate priors. The authors contribute two related conclusions. The first is that if integer networks are used for the whole model, and the model is trained with integer emulation instead of floating point numbers, the performance of the model degrades since using integer networks limits the approximation capacity. The second, as a remedy for the first, is the suggestion to use integer networks only in the prior calculation. This way, it is possible to preserve rate-distortion performance and enable deterministic calculation. They demonstrate results for a learned image compression model consisting of a variational autoencoder (VAE) with a scale and zero mean hyperprior \cite{balle2018variational}. This model is an earlier example of learned image codecs and has a 3-layer hyperdecoder to calculate the scale entropy parameters.

Later, in \cite{sun_learned_2021pcs} Sun \textit{et al.} present their work for quantizing weights and activations to 8-bit fixed point arithmetic in order to reduce complexity and memory bandwidth, and also to ensure cross-platform decodability. In this work, weights are quantized for both the main part and the hyper part whereas activations are only quantized for the hyper part. As a solution to large loss resulting from quantizing the weights of the main part, they suggest a fine-tuning-based coding performance improvement. They also propose an extension to the LUT of scale values, which enables sharing of CDF (cumulative distribution function) LUTs for mean values by restricting quantized latent values to a certain range around mean values. As a baseline, they use the model in \cite{cheng2019deep} with residual connections in the main part, a Gaussian entropy model with mean and scale estimated by a hyperdecoder. 

A more recent work to solve the platform-independent decoding problem with post training quantization of learned image compression is reported in \cite{he_post-training_2022}. As a baseline,  He \textit{et al.} use the image compression network model with Gaussian mixture entropy model (GMM) estimated jointly by context and hyperprior networks \cite{cheng2020learned}. They implement integer arithmetic with a 32-bit accumulator and quantize weights and activations to 8-bit integers  only in entropy estimation subnetworks, which results in 0.35\% BD-rate degradation from the floating point implementation for the Kodak dataset. Also, they propose an idea to speed up the search of CDF LUTs.


In this paper we suggest a solution for the device interoperability problem by converting parameters and activations of entropy subnetworks to 16-bit integers. Our experiments are based on a learned image codec with GMM and context modeling which further uses conditional coding and color separation. Our work differs from previous work in the baseline codec being more complex, having more layers in the entropy part, and using 16-bit for quantization of weights and activations. 
The learned image compression model which we do our experiments with is shown in Section~\ref{baseline_codec} and the quantization method is explained in Section~\ref{sec:quantization_method}. Experimental results are presented in Section~\ref{sec:experiments} and conclusions are given in Section~\ref{discussion}.
\section{Proposed method} \label{sec:method}
\fancyhead[CH]{To be presented at the Picture Coding Symposium (PCS), 7-9 December 2022, San Jos\'{e}, California, USA}

\begin{figure*}[htbp]
\centerline{\includegraphics[width=19cm]{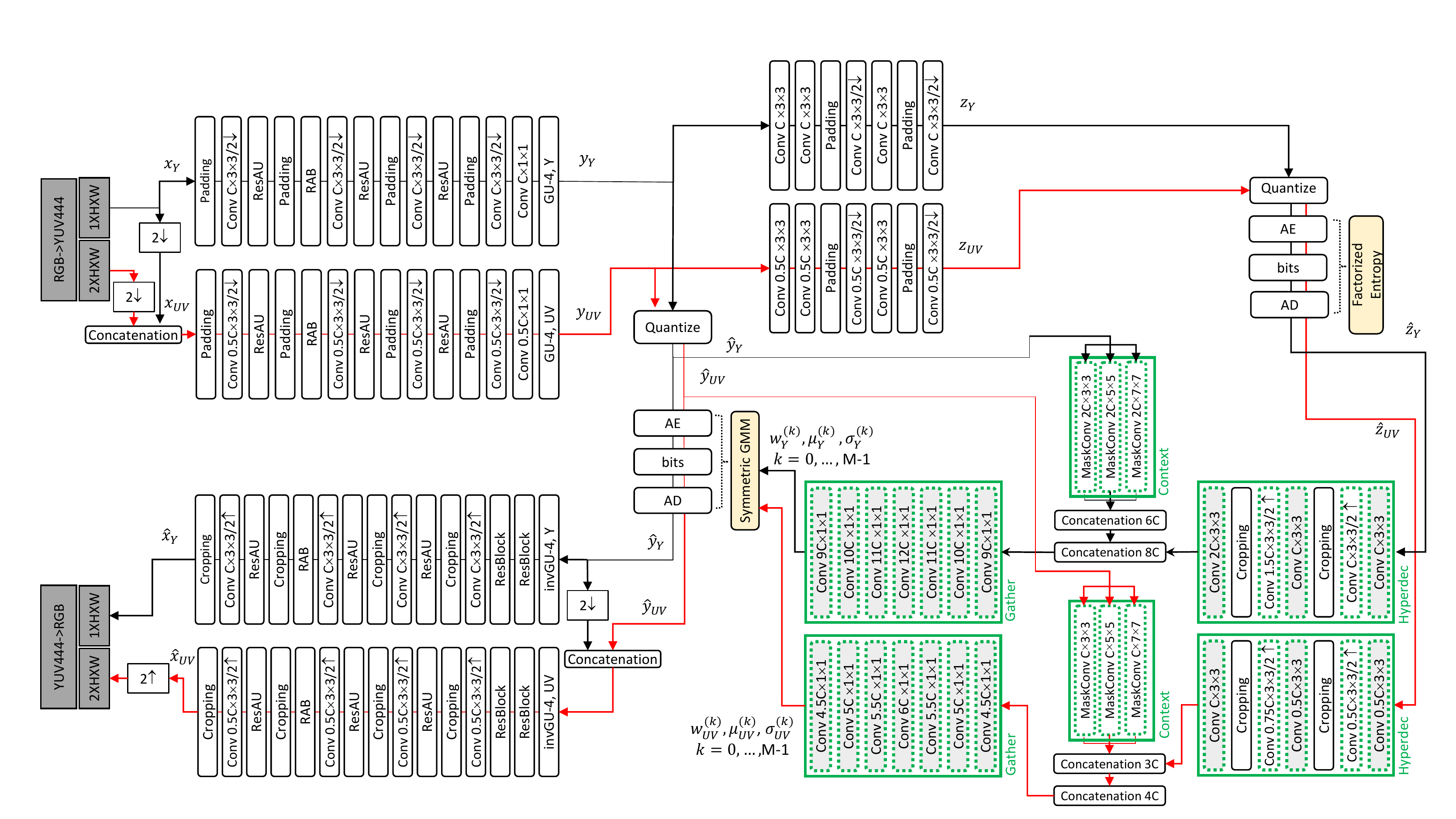}}
\caption{Architecture diagram of baseline codec, C=128.} \color{black}
\label{fig:ccs_overall}
\end{figure*}

We implement quantization to entropy subnetworks of the codec which we experiment with. Our experiments are based on the JPEG AI Call for Proposals (CfP)\footnote[1]{\url{https://jpeg.org/items/20220210_cfp_jpeg_ai.html}} response of TEAM14 \cite{HuaweiCfP} which is a variational autoencoder with color-separated conditional coding, and with context model and GMM entropy model as shown in Fig. \ref{fig:ccs_overall}. The CfP model consists of additional online tools providing rate distortion optimization and rate control functionality, and post filters. However, with the aim of a clean study of the quantization effect, we turn off all tools and filters, i.e., use purely the neural network model in Fig. \ref{fig:ccs_overall} in our experiments. There are four different models trained for four different quality ranges.

\subsection{Description of baseline learned image codec}\label{baseline_codec}
\begin{figure}[htbp]
    \centerline{\includegraphics[width=0.7\columnwidth]{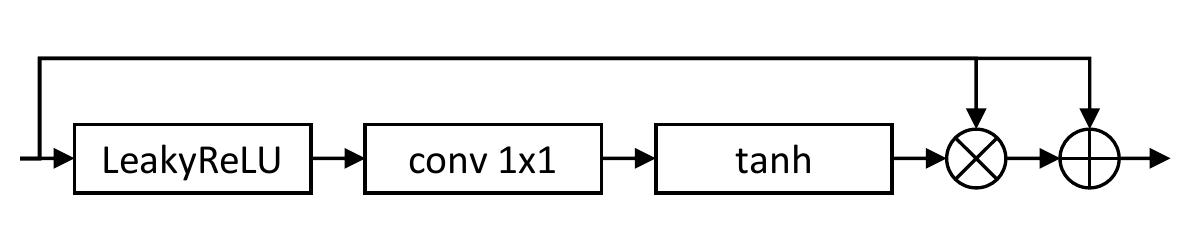}}
    \caption{ResAU module}
    \label{fig:resau}
\end{figure}

\begin{figure}[htbp]
    \centerline{\includegraphics[width=\columnwidth]{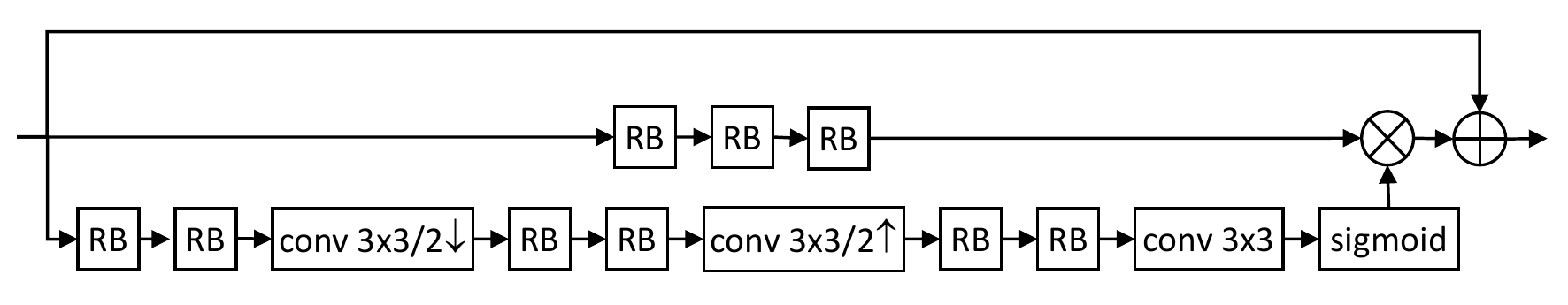}}
    \caption{RAB module}
    \label{fig:rnab}
\end{figure}

\paragraph{Analysis-synthesis part}
In the baseline codec architecture, the luma component (Y) and chroma (U,V) components are processed separately, and the luma component is used as an additional input for chroma encoder and decoder. In the analysis part, there are residual nonlinear units with attention mechanism (ResAU), simplified residual (local) attention blocks (RAB) and gain units (GU). In Fig. \ref{fig:resau} the ResAU module is shown in detail. In Fig. \ref{fig:rnab} the RAB module, residual attention module with trunk and mask branches, is shown, which is adopted from \cite{zhang2019residual}. GU provides variable rate functionality and is implemented as in \cite{gvae}. 

\paragraph{Entropy model of codec}
Entropy parameters estimation is based on a hyperprior and context model. The gather subnetwork outputs the mean, scale and weight of each Gaussian component. We use a GMM with three Gaussian components in the entropy model. Per color component $c=\{Y, U, V\}$ we can write the probability model as in \eqref{eqn:GMM}.
\begin{equation} \label{eqn:GMM}
    p_{\boldsymbol{{\hat{y}}_c}|\boldsymbol{\hat{z}_c}}(\boldsymbol{{{\hat{y}}_c}|\boldsymbol{{\hat{z}}_c}}) \sim \sum_{k=1}^{3}{\boldsymbol{w_c}^{(k)}\mathcal{N}\left(\boldsymbol{\mu_c}^{(k)}, \boldsymbol{\sigma_c}^{2(k)} \right) }
\end{equation}

We use LeakyReLU as the activation function for all layers in the entropy part. At the very last layer of the gather subnetwork, where we output the GMM components, the weights, mean and scale values, Softmax is used as the activation function. To simplify integer calculations, we calculate the Softmax function by using the first order linear approximation of exponential terms in the numerator and the denominator.

\subsection{Quantization method}\label{sec:quantization_method}
We quantize neural network (NN) parameters, input and activations of the neural network layers for subnetworks in the entropy part, i.e, hyper decoder, context, and gather subnetwork, which are shown as the green boxes in Fig. \ref{fig:ccs_overall}. 

The main equation for quantization to a $b$-bit integer with a restricted range is written in \eqref{eqn:int_main}, where $s$ is the scaling factor parameter, clamp$(x; X_1, X_2)$ is the function clipping the $x$ value to the range $[X_1, X_2]$ and $\lfloor \cdot \rceil$ is rounding to the nearest integer value. The accuracy of the quantization depends on the choice of the parameters. In \cite{Nagel_quantization_white}, authors emphasize the trade-off between clipping and rounding error such that one can increase the clipping range in order to decrease clipping error, but including more outliers requires a larger scaling factor, and thus results in a larger rounding error. 
\begin{equation} \label{eqn:int_main}
    {x}_q= \mathrm{clamp}\left(\left\lfloor\frac {x}{s}\right\rceil; -(2^{b-1}-1), 2^{b-1}  - 1\right)
\end{equation}

Since we convert neural network parameters and layer inputs to 16-bit integers, calculate convolutions with integer values and use 32-bit register to store results, there is a risk of overflow. To make sure there is no overflow, we calculate the scaling factors of weights quantization according to weights data and input clipping values. We first decide on the input bit depth, the clipping range for input of each layer, and then calculate the weights scaling factor based on the accumulator size and the bit depth of the input register. The decision of input clipping thresholds is made empirically, searching the best by evaluating the model on calibration data, i.e., a few images, and choosing the best parameter combination per layer to minimize the bit rate. Details of quantization for activations and NN parameters are presented below in this section.

We denote the weight parameters of the NN with $\mathbf{W} \in \mathbb{R}^{m\times K \times K \times n}$, where $m$ and $n$ are the number of input and output channels, respectively, and $K$ is the kernel size of the convolution layer. $\mathbf{x} \in \mathbb{R}^{m\times h \times w}$ is the input of a layer and $\mathbf{b} \in \mathbb{R}^{n}$ is the bias parameter.
\paragraph{Activation quantization}
The main equation for quantizing the input of a layer, $\mathbf{x}$, is given in \eqref{eqn:quant_activation}. 
\begin{equation} \label{eqn:quant_activation}
    \mathbf{x_{q}}= \mathrm{clamp}\left(\left\lfloor\frac {\mathbf{x}}{ s_{\mathbf{x}}}\right\rceil; -(2^{N_{I}-1}-1), 2^{N_{I}-1}  - 1\right)
\end{equation}
We use scaling factors that are a power of two, so we decide on the shift value, $p$, per layer which is the corresponding power of two for the scaling factor with the relation in \eqref{eqn:s_p_relation}.
\begin{equation} \label{eqn:s_p_relation}
    s_{\mathbf{x}}=2^{-p}
\end{equation}

We search quantization parameters by evaluating the model on calibration data. The shift values for the scaling factors are  
\begin{itemize}
    \item $p=8$ for all layers of context, all layers of hyper decoder and the first two layers of gather, 
    \item $p=10$ for the layers 3-7 of the gather network. 
\end{itemize}
The input bit depth $N_I$ values are 
\begin{itemize}
    \item $N_I = 9$ for all layers of context,
    \item  $N_I = 16$ for all layers of hyper decoder and gather.
\end{itemize}

We fuse the scaling of the output of one layer and the input of the next layer into one operation and execute the operations in \eqref{eqn:act_1}-\eqref{eqn:scale_x} in order to calculate the quantized output of convolutions, ${\mathbf{A}}_{\mathbf{q},j}$ .
\begin{equation} \label{eqn:act_1}
    \mathbf{{x^{'}}}= \mathrm{clamp}\left(\left\lfloor{\mathbf{x}}\right\rceil; -2^{N_{I}-1}-1, 2^{N_{I}-1}  - 1\right)
\end{equation}
\begin{equation}
    \mathbf{A^{'}} = \mathbf{b_q} + \mathbf{W_q x'}
    \label{eqn:act_2}
\end{equation}
\begin{equation}
    {\mathbf{A}}_{\mathbf{q},j} = {\frac{\lfloor {\mathbf{A}_{j}^{'}} \rceil}{s_{\mathbf{W},j} {s_{\mathbf{x}}^{'}}}}
    \label{eqn:act_3}
\end{equation}
\begin{equation}
    {s_{\mathbf{x}}}^{'(l)} = 2^{-(p^{(l)}-p^{(l-1)})}
    \label{eqn:scale_x}
\end{equation}

\paragraph{Quantization of NN parameters}
Quantization of weights and bias are executed by scaling and rounding to the nearest integer as written in \eqref{eqn:main_quantization_weights} and \eqref{eqn:bias_main}. We do not use clipping for the weights or bias quantization since the scaling factor is calculated to be the maximum possible for the given weight data. Then, we calculate convolutions with the quantized parameters $\mathbf{W_q}$ and $\mathbf{b_q}$.  
\begin{equation} \label{eqn:main_quantization_weights}
\mathbf{W}_{\mathbf{q}, j,:} = \left\lfloor \frac{\mathbf{W}_{j,:}}{{s_{\mathbf{W},j}}}  \right \rceil
\end{equation}
\begin{equation} \label{eqn:bias_main}
{\mathbf{b}_{\mathbf{q},j}} = \left \lfloor \frac{{\mathbf{b}_j}}{{s_{\mathbf{W},j}} s_{{\mathbf{x}}}} \right \rceil
\end{equation}
After deciding on the input bit depth, $N_I$, per layer, the per-channel scale factor for weights, $s_{\mathbf{W},j}$ for channel $j$, are to be calculated. We use power of two scale factors; hence, there is a corresponding shift value, $k_j$, for each scaling factor $s_{\mathbf{W},j}$ with the relation in \eqref{eqn:weights_per_chanel_scaler}.
\begin{equation} \label{eqn:weights_per_chanel_scaler}
{s_{\mathbf{W},j}} = 2^{-k_j}
\end{equation}
We consider the input bit depth and the accumulator size to calculate a maximum shift value which allows accumulation without overflow. The input bit depth determines the clipping threshold for an input of a layer, as in \eqref{eqn:x_clip_limints} $\forall i$.
\begin{equation} \label{eqn:x_clip_limints}
-2^{N_{I}-1}-1 \leq\mathbf{x}^{'}_i \leq 2^{N_{I}-1}  - 1
\end{equation}
For each output channel $j$, the result of the convolution of quantized input and parameters should be within the limits which are determined by the accumulator register size, $N_A=32$, so that there is no overflow. This is shown in \eqref{eqn:conv_limits_accumulator}.
\begin{equation} \label{eqn:conv_limits_accumulator}
   -2^{N_{A}-1} \leq \sum_{i=1} ^ {m}{\mathbf{W}_{\mathbf{q},{ij}} \mathbf{x}^{'}_i + \mathbf{b}_{\mathbf{q},{j}}} \leq 2^{N_{A}-1}  - 1
\end{equation}
The bias term  is added to each output channel only after multiplications of all input channels, and weights are summed up, so the bias term can be considered at the very end. Combining restrictions from \eqref{eqn:x_clip_limints} and \eqref{eqn:conv_limits_accumulator}, we can write \eqref{eqn:abs_weights_limit} for the case of zero bias. 
\begin{equation} \label{eqn:abs_weights_limit}
   \sum_{i=1} ^ {m} {|\mathbf{W}_{\mathbf{q},{ij}}| \leq 2^{N_{A}-N_{I}}} 
\end{equation}
If we replace $\mathbf{W}_{\mathbf{q}}$ with \eqref{eqn:main_quantization_weights}, we can see the relation with the scaling factor as in \eqref{eqn:abs_weights_limit2}.
\begin{equation} \label{eqn:abs_weights_limit2}
   \sum_{i=1} ^ {m} {\left|\left\lfloor \frac{\mathbf{W}_{ij}}{{s_{\mathbf{W},j}}}  \right \rceil\right| \leq 2^{N_{A}-N_{I}}} 
\end{equation}
Equation \eqref{eqn:weights_per_chanel_shift} shows the maximum per-channel shift factor, $k_{j}$, to ensure there is no overflow for the accumulator size $N_A$ and the input bit depth $N_I$.
\begin{equation} \label{eqn:weights_per_chanel_shift}
{k_{j}} = N_{A} - N_{I} - \left \lceil \mathrm{log}_2 \left(\sum_{i=1} ^ {m} {|\mathbf{W}_{ij}|}\right)\right\rceil
\end{equation}
If there is a non-zero bias term, we update the per-channel shifts for weights calculated in \eqref{eqn:weights_per_chanel_shift}, and choose the limiting value according to \eqref{eqn:bias_modified_per_chanel_shift}.
\begin{equation}\label{eqn:bias_modified_per_chanel_shift}
    {k_j} \xleftarrow{}{\mathrm{min}(N_A-1-p-\mathrm{max}(\lceil \mathrm{log}_2|\mathbf{b}_{j}|\rceil, 0), k_j) - 1}
\end{equation}
%

\subsection{Comparison with previous methods}
The idea of using integer arithmetic in entropy subnetworks to solve the device interoperability problem, which was proposed by Ball\'{e} \textit{et al.} in \cite{balle2018integer}, was later further used in \cite{sun_learned_2021pcs}, \cite{he_post-training_2022} and also is common to our work. As Ball\'{e} \textit{et al.} point out, since encoder and decoder share these parts of the network, they should be calculated the same in the encoder and decoder processors\cite{balle2018integer}. 

One difference from the previous work in ours is the baseline codec, particularly the architecture of the entropy subnetworks of the baseline codec. In terms of codec architecture, compared to \cite{balle2018integer} and \cite{sun_learned_2021pcs}, there is the GMM entropy modeling with joint autoregressive and hyper priors in our baseline. Compared to \cite{he_post-training_2022} which also use joint autoregressive priors and GMM entropy model, the number of layers in each entropy subnetworks in our baseline model is larger. This increases the risk of accumulation of error resulting from the quantization of network parameters and layer inputs.

We also compare the details of quantization implementation with the previous related work. In our implementation we choose 16-bit integers for weights and activations and an accumulator size of 32-bit whereas \cite{balle2018integer}, \cite{sun_learned_2021pcs} and \cite{he_post-training_2022} implement the quantization of weights and activations to 8-bit integers with 32-bit accumulators, and in \cite{balle2018integer} it is additionally suggested to implement with 4-bit weight and activation quantization with 16-bit accumulators. Even though quantizing to a lower number of bits can be a more error-inducing operation, and hence a more difficult problem, when both operands of convolution are 16-bit, the risk of overflow of 32-bit register is higher, compared to the case of operands both being 8-bit. Since numeric overflow is processed differently on different platforms, it needs to be eliminated. Our method contributes a guarantee that there is no overflow while doing a lighter quantization to parameters and inputs. 

We propose a 16-bit quantization of entropy networks for a complex image compression network, where we choose weight scaling factors based on the input bit depth to eliminate the overflow of accumulator, and thus, can achieve device interoperability, while keeping the performance of floating point inference.










\section{Experimental Results} \label{sec:experiments}
Our model is implemented in PyTorch. For quantization we overwrite the convolution layers to emulate quantization with rounding and clipping operations. We aim to reduce the BD-rate \cite{bjontegaard2001calculation} loss when we quantize the entropy subnetworks of the codec, and to eliminate the decoding error which occurs during cross-platform coding.
\paragraph{Training} We train our model before integerization and do not update parameters during integerization, i.e., the integerization approach does not require any training steps. For training we use the JPEG AI training set \cite{JPEGAI_datasets}. 

\paragraph{Evaluation} We evaluate the model before and after quantization, and show that the performance drop is negligible and that there is no decoder error at any of the test images. We evaluate the performance on the Kodak dataset \cite{Kodak_dataset} and Tecnick dataset \cite{asuni2013testimages}, which consists of 100 images with a size of 1200x1200. 
We calculate bit rate, and distortion as YUV-PSNR\footnote[1]{https://gitlab.com/wg1/jpeg-ai/jpeg-ai-qaf}, and also as RGB-PSNR to follow the literature, to evaluate the performance of the models.

Before integerization, we can observe failed reconstructed images as illustrated in Fig. \ref{fig:rec_image}. With our implementation and testing on different devices as listed in Table  \ref{table:device_tests}, we report that for all cases decoding is successful, i.e., there is no failure in any of the reconstructed images, and the BD-rates between encoder and decoder, which are calculated from encoder and decoder reconstructed images, are 0.0\% for all test images.

\begin{figure}[th]
\centerline{\includegraphics[height=4.7cm,trim={0.6cm 0.2cm 1.2cm 1.2cm},clip]{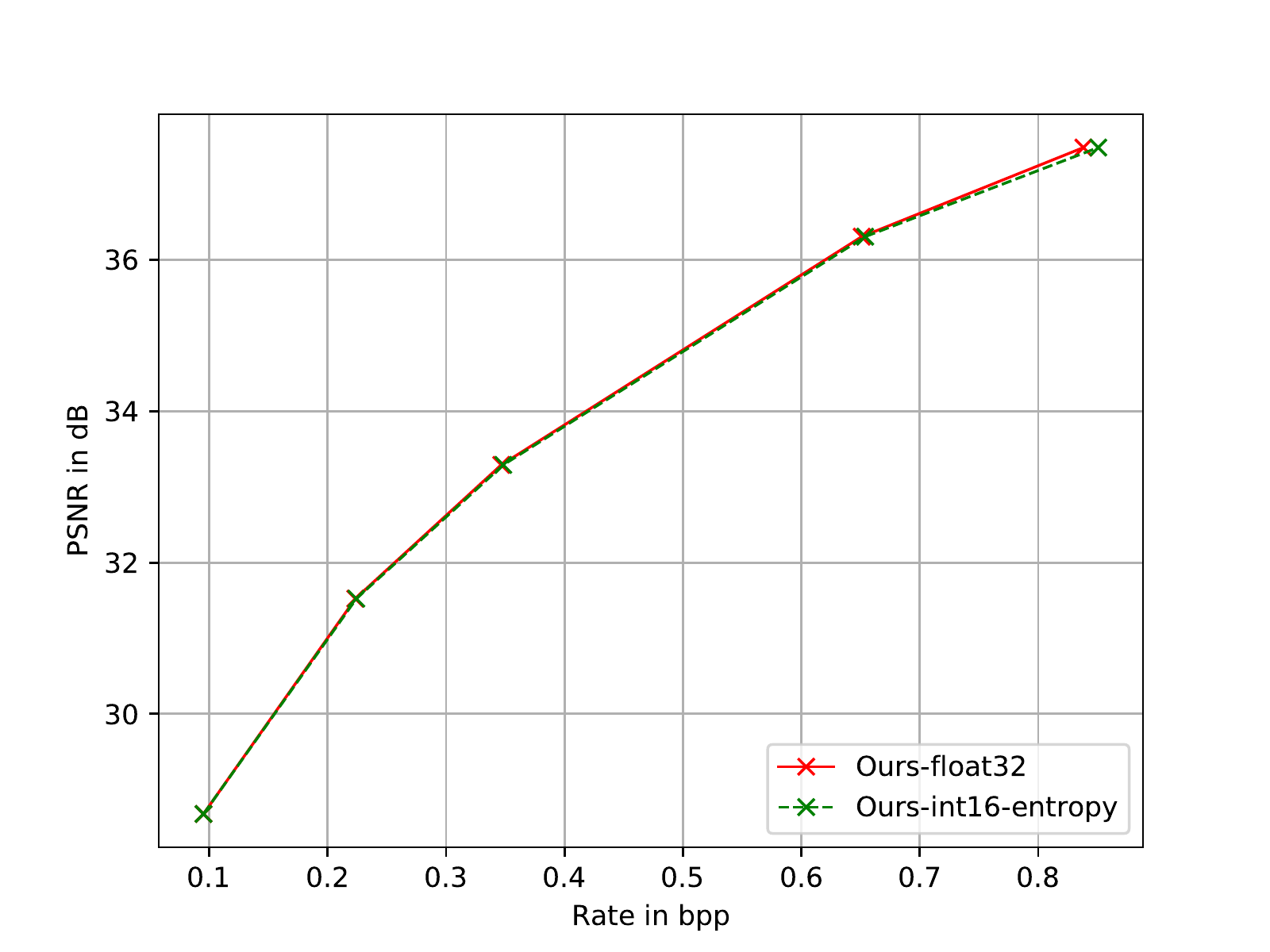}}
\caption{Rate distortion performance on Kodak data set}
\label{fig:psnr_kodak}
\end{figure}

We test our model for both float and integer-emulated versions. Table \ref{table:bdrate_avgkodak} shows the BD-rate results, and Fig. \ref{fig:psnr_kodak} shows the corresponding rate distortion curves for Kodak dataset, with Y-PSNR as the distortion metric. 

In the previous related work, Ball\'{e} \textit{et al.}, Sun \textit{et al.} and He \textit{et al.} all report that with their method of quantizing entropy subnetworks, they can achieve error-free decoding without significant performance degradation from floating point implementation. Following how the success of the previous work is evaluated, we also show that our method does not introduce a significant loss while providing device interoperability.
\begin{table}[]
\addtolength{\tabcolsep}{-1pt}
\caption{Tests for device interoperability}
\begin{center}
\begin{tabular}{|l|l|}
\hline
\textbf{Encoder device}                & \textbf{Decoder device} \\ \hline
\multirow{2}{*}{NVIDIA Titan RTX}      & NVIDIA Titan RTX        \\ \cline{2-2} 
                                       & Intel Core i9-10980XE   \\ \hline
\multirow{2}{*}{NVIDIA V100}           & NVIDIA Titan RTX        \\ \cline{2-2} 
                                       & Intel Core i9-10980XE   \\ \hline
\multirow{2}{*}{Intel Core i9-10980XE} & NVIDIA Titan RTX        \\ \cline{2-2} 
                                       & Intel Core i9-10980XE   \\ \hline
\end{tabular}
\end{center}
\label{table:device_tests}
\end{table}
\begin{table}[]
\addtolength{\tabcolsep}{-1pt}
\caption{BD-rate of integer entropy model vs floating point model}
\begin{center}
\begin{tabular}{|c|c|c|c|c|}
\hline
\textbf{Dataset} & \textbf{\textit{Y-PSNR}} & \textbf{\textit{U-PSNR}} &  \textbf{\textit{V-PSNR}} & \textbf{\textit{RGB-PSNR}}\\
\hline
Average on Kodak & 0.29 & 0.32 & 0.31 & 0.29 \\ 
Average on Tecnick & 0.26 &	0.25 & 0.25 & 0.25 \\ 
\hline
\end{tabular}
\end{center}
\label{table:bdrate_avgkodak}
\end{table}
\section{Discussion} \label{discussion} 

In this paper, we quantize the entropy part of a learned image codec with a hyper-prior, GMM, context models with three different kernel sizes and a seven-layer gather, and achieve deterministic encoding and decoding on different devices with negligible loss compared to the floating point model despite the complexity of the entropy subnetworks.
We propose a method for choosing the quantizing parameters for 16-bit weights and activations quantization such that the coding performance remains the same and overflow is eliminated.

In future work, we plan to experiment with stronger quantization, quantization of weights and activations to a smaller number of bits, and concentrate on actual device implementation issues in order to speed up the decoding and to enable decoding on devices with low computational capacity.

\bibliographystyle{ieeetr}
\bibliography{refs}

\begin{thebibliography}{10}

\bibitem{hevc_2012overview}
G.~J. Sullivan, J.-R. Ohm, W.-J. Han, and T.~Wiegand, ``Overview of the high
  efficiency video coding (\uppercase{HEVC}) standard,'' {\em IEEE Trans.
  Circuits Syst. Video Technol.}, vol.~22, no.~12, pp.~1649--1668, 2012.

\bibitem{vvc}
B.~Bross, Y.-K. Wang, Y.~Ye, S.~Liu, J.~Chen, G.~J. Sullivan, and J.-R. Ohm,
  ``Overview of the versatile video coding (\uppercase{VVC}) standard and its
  applications,'' {\em IEEE Trans. Circuits Syst. Video Technol.}, vol.~31,
  no.~10, pp.~3736--3764, 2021.

\bibitem{balle2018variational}
J.~Ball{\'e}, D.~Minnen, S.~Singh, S.~J. Hwang, and N.~Johnston, ``Variational
  image compression with a scale hyperprior,'' in {\em Proc. Int. Conf. on
  Learning Representations}, 2018.

\bibitem{cheng2020learned}
Z.~Cheng, H.~Sun, M.~Takeuchi, and J.~Katto, ``Learned image compression with
  discretized \uppercase{G}aussian mixture likelihoods and attention modules,''
  in {\em Proc. of the IEEE/CVF Conf. on Comput. Vis. Pattern Recog.},
  pp.~7939--7948, 2020.

\bibitem{minnen2018joint}
D.~Minnen, J.~Ball{\'e}, and G.~D. Toderici, ``Joint autoregressive and
  hierarchical priors for learned image compression,'' {\em Advances in neural
  information processing systems}, vol.~31, 2018.

\bibitem{balle2018integer}
J.~Ball{\'e}, N.~Johnston, and D.~Minnen, ``Integer networks for data
  compression with latent-variable models,'' in {\em Proc. Int. Conf. on
  Learning Representations}, 2018.

\bibitem{sun_learned_2021pcs}
H.~Sun, L.~Yu, and J.~Katto, ``Learned image compression with fixed-point
  arithmetic,'' in {\em 2021 {Picture} {Coding} {Symposium} ({PCS})}, (Bristol,
  United Kingdom), pp.~1--5, June 2021.

\bibitem{cheng2019deep}
Z.~Cheng, H.~Sun, M.~Takeuchi, and J.~Katto, ``Deep residual learning for image
  compression.,'' in {\em CVPR Workshops}, 2019.

\bibitem{he_post-training_2022}
D.~He, Z.~Yang, Y.~Chen, Q.~Zhang, H.~Qin, and Y.~Wang, ``Post-training
  quantization for cross-platform learned image compression,'' {\em arXiv
  preprint arXiv:2202.07513}, 2022.

\bibitem{HuaweiCfP}
E.~Alshina, A.~Boev, Z.~Cui, Y.~Feng, G.~Gaikov, T.~Guo, P.~Jia, A.~Karabutov,
  A.~B. Koyuncu, E.~Koyuncu, M.~Li, J.~Mao, S.~Qian, J.~Sauer, Y.~Shi,
  T.~Solovyev, M.~Sychev, J.~Wang, D.~Yu, and Y.~Zhao, ``Presentation of the
  {Huawei} response to the \uppercase{JPEG AI} {Call for Proposals}: Device
  agnostic learnable image coding using primary component extraction and
  conditional coding,'' Tech. Rep. wg1m96016-REQ, Huawei Technologies, 2022.

\bibitem{zhang2019residual}
Y.~Zhang, K.~Li, B.~Zhong, and Y.~Fu, ``Residual non-local attention networks
  for image restoration,'' in {\em Int. Conf. on Learning Representations},
  2019.

\bibitem{gvae}
Z.~Cui, J.~Wang, B.~Bai, T.~Guo, and Y.~Feng, ``{G-VAE}: A continuously
  variable rate deep image compression framework,'' {\em arXiv preprint
  arXiv:2003.02012}, 2020.

\bibitem{Nagel_quantization_white}
M.~Nagel, M.~Fournarakis, R.~A. Amjad, Y.~Bondarenko, M.~van Baalen, and
  T.~Blankevoort, ``A white paper on neural network quantization,'' {\em arXiv
  preprint arXiv:2106.08295}, 2021.

\bibitem{bjontegaard2001calculation}
G.~Bj\o{}ntegaard, ``Calculation of average \uppercase{PSNR} differences
  between \uppercase{RD}-curves,'' {\em VCEG-M33}, 2001.

\bibitem{JPEGAI_datasets}
``J{PEG AI} dataset,'' 2022.
\newblock [Online]. Available "\url{https://jpeg.org/jpegai/dataset.html}".

\bibitem{Kodak_dataset}
E.~Kodak, ``Kodak lossless true color image suite ({P}hoto{CD PCD0992}),''
  1993.
\newblock [Online]. Available "\url{http://r0k.us/graphics/kodak/}".

\bibitem{asuni2013testimages}
N.~Asuni and A.~Giachetti, ``{TESTIMAGES}: A large data archive for display and
  algorithm testing,'' {\em Journal of Graphics Tools}, vol.~17, no.~4,
  pp.~113--125, 2013.

\end{thebibliography}
\end{document}